\documentclass[reqno,12pt]{article}
\usepackage[english]{babel}
\usepackage{cite}

\usepackage{ae} 
\usepackage[T1]{fontenc}
\usepackage{amsmath}
\numberwithin{equation}{section}
\usepackage{enumitem}
\usepackage{amssymb}
\usepackage{mathrsfs}
\usepackage{graphicx}
\usepackage{mdframed}
\usepackage{amsfonts,mathtools}
\usepackage{color}
\definecolor{darkblue}{cmyk}{0.9,0.9,0,0}
\usepackage[colorlinks=true,linkcolor=darkblue,citecolor=darkblue,urlcolor=darkblue]{hyperref}
\usepackage{epsfig}

\usepackage{graphicx}
\usepackage{tikz}

\usepackage{float}
\usepackage{tikz}
\usetikzlibrary{decorations.markings}
\usepackage{tikz,lipsum,lmodern}
\usepackage[most]{tcolorbox}
\usepackage{hyperref}
\usepackage{xcolor}

\def\XXint#1#2#3{{\setbox0=\hbox{$#1{#2#3}{\int}$}
\vcenter{\hbox{$#2#3$}}\kern-.5\wd0}}

\newcommand{\eg}{{\it e.g.,\ }}

\usepackage{varioref}
\usepackage{makeidx}
\makeindex

\newcommand\blfootnote[1]{%
  \begingroup
  \renewcommand\thefootnote{}\footnote{\hspace{-6mm}#1}%
  \addtocounter{footnote}{-1}%
  \endgroup
}

\begin{document}

\thispagestyle{empty}

\renewcommand{\thefootnote}{\fnsymbol{footnote}}
\setcounter{page}{1}
\setcounter{footnote}{0}
\setcounter{figure}{0}

\vspace{-0.4in}

\begin{center}
$$~$$
{\Large
$n-\overline{n}$ Oscillation in $S^1/Z_2\times Z_2'$ Orbifold $SU(5)$ GUT
\par}
\vspace{1.0cm}

{Ankit Das$^\text{\tiny 1, \tiny 2}$, Sarthak Duary$^\text{\tiny 3}$, and Utpal Sarkar$^\text{\tiny 1}$}
\blfootnote{\tt{ankit.d.a.s.20010206@gmail.com, sarthakduary@tsinghua.edu.cn, utpal.sarkar.prl@gmail.com}}
\\ \vspace{1.2cm}
\footnotesize{\textit{
$^\text{\tiny 1}$Indian Institute of Science Education and Research Kolkata, Mohanpur 741246, India \\
$^\text{\tiny 2}$Indian Institute of Technology Kanpur, Kanpur 208016, India \\
$^\text{\tiny 3}$Yau Mathematical Sciences Center (YMSC), Tsinghua University, Beijing 100084, China\\
}
\vspace{4mm}
}
\end{center}

\vspace{2mm}
\begin{abstract}
We explore the possibility of $B$ and $B-L$ violating processes, specifically proton decay and neutron-antineutron oscillation, using explicit realization of operators in the $SU(5)$ grand unified theory with an $S^1/Z_2 \times Z_2'$ orbifold space.
\end{abstract}

\newpage

\addtocontents{toc}{\protect\setcounter{tocdepth}{2}}

\setcounter{page}{1}
\renewcommand{\thefootnote}{\arabic{footnote}}
\setcounter{footnote}{0}

{
\tableofcontents
}

\newpage


\section{Introduction}

The concept of grand unified theories (GUTs) that can unify the Standard Model gauge interactions has been a dominant idea in high-energy physics 
\eg see \cite{Georgi:1974sy, Pati:1973uk}. The success of gauge coupling unification in minimal supersymmetric Standard Model (MSSM) extensions of these GUT theories \eg see \cite{Dimopoulos, Dimopoulos:1981zb} has further strengthened support for the GUT framework. However, despite these significant successes, including the generation of small neutrino masses, the GUT concept is not without its limitations. Some of the key issues include
\begin{enumerate}[label=\roman*.]
\item Problems with the Higgs sector of the high-scale GUT theory, particularly the challenge of achieving the necessary doublet-triplet splitting.
\item The problem of proton decay occurring too rapidly in many GUT models.
\item The mismatch between the GUT scale and the expected scale of unification with gravity.
\end{enumerate}
While the GUT has had significant impact and successes, unresolved issues and challenges persist in constructing a high-energy unification framework. A new avenue of GUT has been developed \eg see \cite{Kawamura:2000ev, Altarelli:2001qj, Hall:2001pg}. The central idea is that the GUT gauge symmetry exists in $5$ dimensions, but is broken down to the Standard Model (SM) symmetry by imposing GUT-violating boundary conditions on a singular orbifold compactification. Given the successful predictions of supersymmetric gauge coupling unification in traditional GUT models, the most appealing class of models incorporates both supersymmetry and an $SU(5)$ gauge symmetry in $5$ dimensions. In this scenario, the compactification on the orbifold $S^1/Z_2\times Z_2'$ breaks both the higher-dimensional GUT group and $5d$ supersymmetry, resulting in a $4$-dimensional $\mathcal{N}=1$ supersymmetric model with the SM gauge group. In orbifold GUTs, tree-level processes mediating proton decay are precisely projected out by the orbifold symmetry breaking. Proton decay can still occur, but only through higher-order suppressed operators. Judicious choices of matter parity assignments can even completely forbid proton decay to all orders. 

We explore the phenomenon of tree level proton decay within the framework of minimal supersymmetric $SU(5)$ GUT that include an additional compact spatial dimension, represented by the orbifold $S^1/Z_2\times Z_2'$. We find that in the $5$-dimensional spacetime with the $5$th dimension compactified on the $S^1/Z_2 \times Z'_2$ orbifold, and with a supersymmetric $SU(5)$ gauge symmetry, the issues of triplet-doublet splitting and suppression of tree-level proton decay processes are naturally realized. In this model, the absence of tree-level proton decay seems intrinsically linked to the orbifold construction. Also, by including Higgs fields transforming as $\textbf{10}$ and $\textbf{15}$ representations of $SU(5)$, our model allows for neutron-antineutron oscillations ($n-\bar{n}$-oscillations), which could potentially play a role in baryogenesis. Furthermore, these additional $\textbf{10}$ and $\textbf{15}$ Higgs representations can even generate light neutrino masses in a non-supersymmetric SU(5) theory compactified on the $S^1/Z_2 \times Z'_2$ orbifold.
\subsection*{Organization of the paper.}
This paper is organized as follows. In section \ref{review}, we give the necessary preliminaries \eg $SU(N)$ group representation, $SU(5)$ unification of particles, and review of $S^1/Z_2\times Z_2'$ orbifold. Next, in section \ref{grand unification}, we study grand unification in $S^1/ Z_2 \times Z_2'$ orbifold as a prepartion for exploring proton decay, and $n-\Bar{n}$ in this model. In section \ref{no pdecay}, we study no proton decay in orbifold $SU(5)$ GUT. Next, in section \ref{nnbar}, we study $n-\Bar{n}$ oscillation in orbifold $SU(5)$ GUT. Finally, we conclude in section \ref{conclusions}.

\section{Preliminaries}
In this section we give the necessary preliminaries to make our paper self-contained. We begin with subsection \ref{SU(N)}, where we describe $SU(N)$ group representation. Then we describe $SU(5)$ unification of particles in subsection \ref{unif}. In subsection \ref{review}, we review the $S^1/Z_2\times Z_2'$ orbifold.

We consider $SU(5)$ group, which comprises special unitary transformations that operate on five complex variables. Particle multiplets organize into group representations, which implies that symmetry groups interact with and rearrange them. To effectively categorize particle multiplets, we must first categorize these group representations. In the subsection \ref{SU(N)}, we will describe $SU(N)$ group representation. 

\subsection{$SU(N)$ group representations}
\label{SU(N)}
The defining representation for the $SU(N)$ group is $N$-dimensional, and the special unitary matrices are $(N \times N)$-matrices. The $N$ and $\bar{N}$ are both $N$-dimensional representations. Here, $\bar{N}$ is the complex-conjugate of the defining representation. We can assign labels to single-particle states as $\psi_i$ and to two-particle states as $\psi_{ij} \to \psi_i \varphi_j$. In this context, $\psi_i$ ranges from $1$ to $N$, while for $\psi_{ij}$, there are $N^2$ potential entries. We also have two distinct representations, $\psi_{ij}(\text{symmetric})$, and $\psi_{ij}(\text{antisymmetric})$. We take an example for two spin states, which are labeled as either ``up'' or ``down''. In total, there are four states: $uu, ud, du$, and $dd$ representing spins pointing either up or down. These states pertain to two spin-$\frac{1}{2}$ particles. Combining two such spin-$\frac{1}{2}$  particles results in four possible states, but these states form multiplets among themselves. One of these combinations is a singlet, corresponding to a spin of $0$. The other combination is a triplet of states, resulting in a spin of $1$. The linearly independent combinations that are symmetric with respect to the interchange of $i$ and $j$, i.e., when $\psi_{ij}=\psi_{ji}$, are as follows 
\begin{equation}
\begin{split}
uu, dd, \frac{1}{\sqrt{2}}(ud+du).
\end{split}
\end{equation} 
These combinations exhibit the property that when acted upon by the $SU(2)$ group, they mix with each other but do not mix with the antisymmetric combination $\frac{1}{\sqrt{2}}(ud-du)$. The singlet, represented as $\frac{1}{\sqrt{2}}(ud-du)$ remains unchanged under group rotations. Consequently, when combining two spin-$\frac{1}{2}$ particles, one combination yields a spin of $1$, while the other yields a spin of $0$.

\subsection{$SU(5)$ unification of particles}
\label{unif}
We are focusing on the $SU(5)$ group, $SU(5)$ mixes up $5$-fermions belonging to a single generation. For the defining representation, the $5$-entries are  
\begin{equation}
\label{multiplet}
\text{[Multiplet under the group SU(5)]}_L\equiv5=	
\begin{pmatrix}
	\nu_e\\
	e^{-}\\  
	d^c\\
	d^c\\
	d^c
\end{pmatrix}
.
\end{equation}
where, $\nu_e, 	e^{-}, d^c$ are the neutrino, the electron, and the down antiquark. \footnote{We treat the particles as left-handed. This is simply a way of enumerating the particles as left-handed representation of $SU(5)$. It serves a purpose because the $SU(5)$ group does not convert left-handed particles into right-handed particles and vice versa. Note that, the anti-particle of the left-handed positron $e^{+}_L$ is the right-handed electron, $e^{-}_R$.} The $SU(2)$ of the weak interactions mix neutrino with electron, and the color group transformation of quantum chromodynamics, $SU(3)$ mix different colors (red(R), green(G), and blue(B)). However, there are new possibilities, such as transformations that convert $\nu_e$ into $d^c$ quarks or $e^{-}$ into $d^c$ quarks, and so forth. For each of these transformations, there typically exists a gauge boson that facilitates transitions between them. The gauge bosons responsible for converting one $d^c$ quark into another are the gluons, associated with strong interaction and described by $SU(3)$. Meanwhile, the gauge bosons responsible for interchanging electrons and neutrinos are the $W, Z$ bosons, and photons ($\gamma$), which are related to weak interaction and governed by $SU(2)$. Now, let us  move to the transitions which can happen between these particles by virtue of emission, and absorption of gauge bosons. The $SU(5)$ group is bigger than the $SU(3) \times SU(2) \times U(1)$, leading to the presence of novel types of gauge bosons. However, if we attempt to unify $SU(3) \times SU(2) \times U(1)$ into $SU(5)$, new varieties of gauge bosons must necessarily exist, specifically those that enable transitions from $d^c$ to an electron $e^{-}$.

 The quantum numbers of paticles in a theory like this are related to the generators of the group. The generators of the group are the $N \times N$ traceless hermitian matrices, which in fact are in one-to-one correspondence with the adjoint representation. The number of independent traceless hermitian matrices of dimension $N$ are $N^2-1$. The generators of the group are intimately connected with gauge boson emission. One of the generators of the group has to be associated with the quantum mechanical operator which represents electric charge. The requirement that the determinant equals one for unitary matrices implies that the trace must be zero for these generators. Now, the trace of the generator is zero implies sums of the charge within a given representation always add up to zero. In eq.\eqref{multiplet}, we see that to embed electric charge  into the generators of $SU(5)$, the electric charge is the multiplet is $-1+3\times \frac{1}{3}=0.$

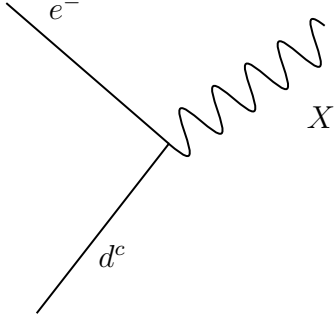
\begin{figure}[H]
	\centering
  \tikzset{every picture/.style={line width=0.75pt}} 

\begin{tikzpicture}[x=0.75pt,y=0.75pt,yscale=-1,xscale=1]
	
	\draw    (118,134) -- (200,205) ;
	\draw    (200,205) -- (133.33,290.56) ;
	\draw   (199.63,204.76) .. controls (204.16,208.52) and (208.48,212.09) .. (209.98,211.07) .. controls (211.47,210.05) and (209.73,204.72) .. (207.9,199.13) .. controls (206.06,193.54) and (204.32,188.21) .. (205.82,187.2) .. controls (207.31,186.18) and (211.63,189.75) .. (216.16,193.5) .. controls (220.69,197.26) and (225.01,200.83) .. (226.51,199.81) .. controls (228,198.79) and (226.26,193.46) .. (224.43,187.87) .. controls (222.59,182.28) and (220.85,176.95) .. (222.35,175.94) .. controls (223.84,174.92) and (228.16,178.49) .. (232.69,182.24) .. controls (237.22,186) and (241.54,189.57) .. (243.04,188.55) .. controls (244.53,187.53) and (242.79,182.2) .. (240.96,176.61) .. controls (239.12,171.02) and (237.38,165.7) .. (238.88,164.68) .. controls (240.37,163.66) and (244.69,167.23) .. (249.22,170.98) .. controls (253.75,174.74) and (258.07,178.31) .. (259.57,177.29) .. controls (261.06,176.27) and (259.32,170.95) .. (257.49,165.35) .. controls (255.65,159.76) and (253.91,154.44) .. (255.41,153.42) .. controls (256.9,152.4) and (261.22,155.97) .. (265.75,159.72) .. controls (270.28,163.48) and (274.6,167.05) .. (276.1,166.03) .. controls (277.59,165.01) and (275.85,159.69) .. (274.02,154.1) .. controls (272.18,148.5) and (270.44,143.18) .. (271.94,142.16) .. controls (273.01,141.43) and (275.55,143.07) .. (278.58,145.45) ;
	
	\draw (138,128.4) node [anchor=north west][inner sep=0.75pt]    {$e^{-}$};
	\draw (163,254.4) node [anchor=north west][inner sep=0.75pt]    {$d^c$};
	\draw (268,184.4) node [anchor=north west][inner sep=0.75pt]    {$X$};

\end{tikzpicture}
	\caption{Potential new kinds of interations: emission of $X$-boson}
	\label{fig:xboson}
\end{figure}

\begin{figure}[H]
	\centering
\tikzset{every picture/.style={line width=0.75pt}} 

\begin{tikzpicture}[x=0.75pt,y=0.75pt,yscale=-1,xscale=1]
	
	\draw    (118,134) -- (200,205) ;
	\draw    (200,205) -- (133.33,290.56) ;
	\draw   (199.63,204.76) .. controls (204.16,208.52) and (208.48,212.09) .. (209.98,211.07) .. controls (211.47,210.05) and (209.73,204.72) .. (207.9,199.13) .. controls (206.06,193.54) and (204.32,188.21) .. (205.82,187.2) .. controls (207.31,186.18) and (211.63,189.75) .. (216.16,193.5) .. controls (220.69,197.26) and (225.01,200.83) .. (226.51,199.81) .. controls (228,198.79) and (226.26,193.46) .. (224.43,187.87) .. controls (222.59,182.28) and (220.85,176.95) .. (222.35,175.94) .. controls (223.84,174.92) and (228.16,178.49) .. (232.69,182.24) .. controls (237.22,186) and (241.54,189.57) .. (243.04,188.55) .. controls (244.53,187.53) and (242.79,182.2) .. (240.96,176.61) .. controls (239.12,171.02) and (237.38,165.7) .. (238.88,164.68) .. controls (240.37,163.66) and (244.69,167.23) .. (249.22,170.98) .. controls (253.75,174.74) and (258.07,178.31) .. (259.57,177.29) .. controls (261.06,176.27) and (259.32,170.95) .. (257.49,165.35) .. controls (255.65,159.76) and (253.91,154.44) .. (255.41,153.42) .. controls (256.9,152.4) and (261.22,155.97) .. (265.75,159.72) .. controls (270.28,163.48) and (274.6,167.05) .. (276.1,166.03) .. controls (277.59,165.01) and (275.85,159.69) .. (274.02,154.1) .. controls (272.18,148.5) and (270.44,143.18) .. (271.94,142.16) .. controls (273.01,141.43) and (275.55,143.07) .. (278.58,145.45) ;
	
	\draw (138,128.4) node [anchor=north west][inner sep=0.75pt]    {$\nu_e$};
	\draw (163,254.4) node [anchor=north west][inner sep=0.75pt]    {$d^c$};
	\draw (268,184.4) node [anchor=north west][inner sep=0.75pt]    {$Y$};

\end{tikzpicture}

	\caption{Potential new kinds of interations: emission of $Y$-boson}
	\label{fig:yboson}
\end{figure}
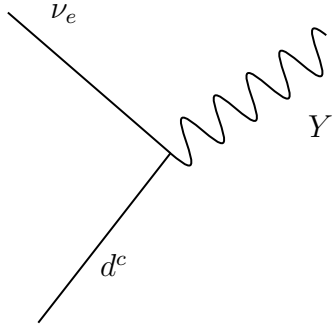

 In figure \ref{fig:xboson}, there is a transition depicted where $d^c$ transforms into an electron ($e^{-}$) while emitting the $X$-boson. In this context, the $X$-boson carries a charge of $\frac{4}{3}$. This charge assignment is consistent with the principle of charge conservation in the process, as the $d^c$ quark has a charge of $\frac{1}{3}$, while the electron ($e^{-}$) has a charge of $-1$.
In figure \ref{fig:yboson}, we observe the emission of the $Y$-boson, which transforms a $d^c$ quark into a $\nu_e$ neutrino. In this context, the $Y$-boson carries a charge of $\frac{1}{3}$.
It is impotant to note that to maintain color conservation, the $X$ and $Y$-bosons must possess color charges. This implies that they have the capability to emit and absorb gluons themselves.

The other fermions we left out can be placed within a different representation.
We take the $(\bar{5} \times \bar{5})$-antisymmetric combination given by 
\begin{equation}
\label{10bar}
(\bar{5} \times \bar{5})_{\text{antisymmteric}}\equiv \overline{10}=
  \begin{pmatrix}
	0 & {e^{+}} & {d} & {d} & {d} \\
	 & 0 & {u} & {u} & {u} \\
	 &  & 0 & {u^c} & u^c \\
	 &  &  & 0 & u^c \\
	 &  &  &  & 0 
\end{pmatrix}
.
\end{equation}
Here, we have 
\begin{equation}
	\bar{5}=	
	\begin{pmatrix}
		\nu^c_e\\
		e^{+}\\  
		{d}\\
		{d}\\
		{d}
	\end{pmatrix}
	.
\end{equation}
The $SU(5)$ unification of particles is given by $5$, and $\overline{10}$. We have the breakdown of particle-multiplets into a $5$, and $\overline{10}$ of $SU(5)$. 

\begin{figure}[H]
	\centering
 \tikzset{every picture/.style={line width=0.75pt}} 

\begin{tikzpicture}[x=0.75pt,y=0.75pt,yscale=-1,xscale=1]
	
	\draw    (118,134) -- (200,205) ;
	\draw    (200,205) -- (129,287) ;
	\draw   (199.27,203.82) .. controls (200.96,207.32) and (202.58,210.66) .. (204.39,210.63) .. controls (206.2,210.59) and (207.7,207.2) .. (209.27,203.64) .. controls (210.83,200.08) and (212.34,196.69) .. (214.14,196.66) .. controls (215.95,196.63) and (217.57,199.96) .. (219.26,203.47) .. controls (220.96,206.97) and (222.57,210.31) .. (224.38,210.28) .. controls (226.19,210.25) and (227.69,206.85) .. (229.26,203.29) .. controls (230.83,199.73) and (232.33,196.34) .. (234.14,196.31) .. controls (235.95,196.28) and (237.57,199.61) .. (239.26,203.12) .. controls (240.95,206.62) and (242.57,209.96) .. (244.38,209.93) .. controls (246.19,209.9) and (247.69,206.51) .. (249.26,202.94) .. controls (250.83,199.38) and (252.33,195.99) .. (254.14,195.96) .. controls (255.95,195.93) and (257.57,199.26) .. (259.26,202.77) .. controls (260.95,206.27) and (262.57,209.61) .. (264.38,209.58) .. controls (266.19,209.55) and (267.69,206.16) .. (269.26,202.59) .. controls (270.83,199.03) and (272.33,195.64) .. (274.14,195.61) .. controls (275.94,195.58) and (277.56,198.92) .. (279.26,202.42) .. controls (280.95,205.93) and (282.57,209.26) .. (284.37,209.23) .. controls (286.18,209.2) and (287.68,205.81) .. (289.25,202.25) .. controls (290.82,198.68) and (292.32,195.29) .. (294.13,195.26) .. controls (295.94,195.23) and (297.56,198.57) .. (299.25,202.07) .. controls (299.37,202.31) and (299.48,202.54) .. (299.59,202.77) ;
	\draw    (299.5,203.5) -- (370.5,121.5) ;
	\draw    (367.93,287.66) -- (299.5,203.5) ;
	\draw    (396,128) -- (396,289) ;
	\draw   (94,325) .. controls (94,313.95) and (166.31,305) .. (255.5,305) .. controls (344.69,305) and (417,313.95) .. (417,325) .. controls (417,336.05) and (344.69,345) .. (255.5,345) .. controls (166.31,345) and (94,336.05) .. (94,325) -- cycle ;
	\draw   (215,353) .. controls (215,357.67) and (217.33,360) .. (222,360) -- (255,360) .. controls (261.67,360) and (265,362.33) .. (265,367) .. controls (265,362.33) and (268.33,360) .. (275,360)(272,360) -- (308,360) .. controls (312.67,360) and (315,357.67) .. (315,353) ;
	\draw   (265.31,367.38) -- (313,368) -- (312.69,391.62) ;
	\draw   (422,118) .. controls (422,113.33) and (419.67,111) .. (415,111) -- (392,111) .. controls (385.33,111) and (382,108.67) .. (382,104) .. controls (382,108.67) and (378.67,111) .. (372,111)(375,111) -- (345,111) .. controls (340.33,111) and (338,113.33) .. (338,118) ;
	\draw    (383,105) -- (441,105) ;
	\draw    (551,194) -- (550,281) ;
	\draw    (479,130) -- (551,194) ;
	\draw    (551,194) .. controls (550.74,191.66) and (551.78,190.36) .. (554.12,190.09) .. controls (556.46,189.82) and (557.5,188.52) .. (557.23,186.18) .. controls (556.97,183.84) and (558.01,182.54) .. (560.35,182.27) .. controls (562.69,182) and (563.73,180.7) .. (563.47,178.36) .. controls (563.2,176.02) and (564.24,174.72) .. (566.58,174.45) .. controls (568.92,174.18) and (569.96,172.88) .. (569.7,170.54) .. controls (569.44,168.2) and (570.48,166.9) .. (572.82,166.63) .. controls (575.16,166.36) and (576.2,165.06) .. (575.93,162.72) .. controls (575.67,160.38) and (576.71,159.08) .. (579.05,158.81) .. controls (581.39,158.54) and (582.43,157.24) .. (582.17,154.9) .. controls (581.9,152.56) and (582.94,151.26) .. (585.28,150.99) .. controls (587.62,150.72) and (588.66,149.42) .. (588.4,147.08) .. controls (588.14,144.74) and (589.18,143.44) .. (591.52,143.17) .. controls (593.86,142.9) and (594.9,141.6) .. (594.63,139.26) .. controls (594.37,136.92) and (595.41,135.62) .. (597.75,135.35) .. controls (600.09,135.08) and (601.13,133.78) .. (600.86,131.44) .. controls (600.6,129.1) and (601.64,127.8) .. (603.98,127.53) -- (606,125) -- (606,125) ;
	
	\draw (138,128.4) node [anchor=north west][inner sep=0.75pt]    {$e^{+}$};
	\draw (163,254.4) node [anchor=north west][inner sep=0.75pt]    {$d$};
	\draw (245,219.4) node [anchor=north west][inner sep=0.75pt]    {$X$};
	\draw (326,252.4) node [anchor=north west][inner sep=0.75pt]    {$u$};
	\draw (354,145.4) node [anchor=north west][inner sep=0.75pt]    {$u^c$};
	\draw (407,257.4) node [anchor=north west][inner sep=0.75pt]    {$u$};
	\draw (409,131.4) node [anchor=north west][inner sep=0.75pt]    {$u$};
	\draw (308,388.4) node [anchor=north west][inner sep=0.75pt]    {$p$};
	\draw (445,94.4) node [anchor=north west][inner sep=0.75pt]    {$\pi ^{0}$};
	\draw (436,196.4) node [anchor=north west][inner sep=0.75pt]    {$\Longrightarrow $};
	\draw (545,287.4) node [anchor=north west][inner sep=0.75pt]    {$p$};
	\draw (481,108.4) node [anchor=north west][inner sep=0.75pt]    {$e^{+}$};
	\draw (607,103.4) node [anchor=north west][inner sep=0.75pt]    {$\pi ^{0}$};

\end{tikzpicture}

	\caption{Proton decay}
	\label{fig:protondecay}
\end{figure}
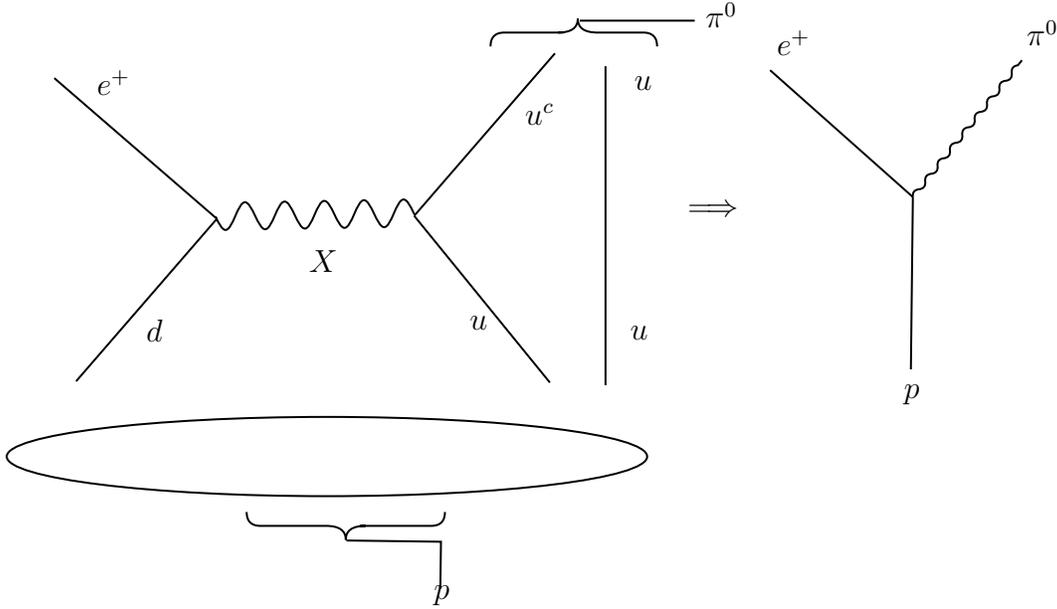

Writing explicitly the colors, we have \footnote{\begin{equation}
		\bar{5}=	
		\begin{pmatrix}
			\nu^c_e\\
			e^{+}\\  
			{d_R}\\
			{d_G}\\
			{d_B}
		\end{pmatrix}
		.
\end{equation}}
\begin{equation}
	\label{10bar}
	(\bar{5} \times \bar{5})_{\text{antisymmteric}}\equiv \overline{10}=
	\begin{pmatrix}
		0 & {e^{+}} & {d_R} & {d_G} & {d_B} \\
		-e^{+}& 0 & {u_R} & {u_G} & {u_B} \\
		-d_R& -u_R & 0 & u^c_B & u^c_G \\
		-d_G&  -u_G&-u^c_B  & 0 & u^c_R \\
		-d_B&-u_B  &  -u^c_G& -u^c_R & 0 
	\end{pmatrix}
	.
\end{equation}
Now, the emission of an $X$-boson can facilitate the transformation of $d^c$ into $e^{-}$. The emission of $X$ results in the transformation of a $d$-quark into an $e^{+}$ (positron). This implies the existence of novel types of processes.
We can have the proton decay shown in figure \ref{fig:protondecay}. 
The $X$ has charge $-\frac{4}{3}$ since $d$ has charge $-\frac{1}{3}$, and $e^{+}$ has charge $+1$. We can have $X$-boson for the process $u +X \to u^c$, in which the charge conservation becomes $-\frac{4}{3}+\frac{2}{3}=-\frac{2}{3}$. Here is two processes that can happen involving $X$. The processes are
\begin{equation}
\begin{split}
&d \to e^{+}+X\\
&u +X\to u^c.
\end{split}
\end{equation}
We get proton becomes $e^{+}$ and $\pi^0$
\begin{equation}
p \to e^{+}+\pi^{0}.
\end{equation}

\subsection{Review of $S^1/Z_2\times Z_2'$ orbifold}
\label{review}
Let us revisit and examine more closely the key characteristics of the $S^1/Z_2\times Z_2'$ orbifold model as described in \cite{Kawamura:2000ev, Altarelli:2001qj, Hall:2001pg}.
We consider $S^1$ which is a circle of radius $R$ where $R^{-1}$ is the grand unification scale $M_{\mathrm{GUT}}$. Now, the orbifold is a type of space that is obtained by dividing a manifold by the action of a discrete transformation group. This process results in a space that has fixed points-these are points that get mapped back to themselves under the discrete transformation.
A simple example of an orbifold is $S^1/Z_2$. This is constructed by taking a circle $S^1$ (with radius $R$) and dividing it by the $Z_2$ transformation. The $Z_2$ transformation identifies points on the circle that are related by a reflection.
The orbifold $S^1/Z_2\times Z_2'$ is obtained by first gauging out the $S^1$ with $Z_2$ transformation, and then again gauging the quotient space by the transformation $Z_2'$ given by 
\begin{align}\label{lol}
	Z_2:& y\rightarrow -y ,\\
	Z_2':& y'\rightarrow -y'; \ \ \ y'=y+\pi R/2.
\end{align}
We have two singular (fixed) points at $y=0$ and $y=\pi R/2$, and thus we will be working on the line segment $[0,\pi R/2]$ (with length $\pi R/2$). Let us consider a brane world scenario that utilizes the fixed points of an orbifold. In this setup, we assume that the spacetime is factorized into a product of
\begin{enumerate}[label=\roman*.]
    \item A $4$-dimensional Minkowski spacetime.
    \item An orbifold C such as $S^1/Z_2$, or $S^1/Z_2\times Z_2'$.
\end{enumerate}
The key idea is that the underlying spacetime geometry is not simply the $4$-dimensional space, but rather a product space that includes the orbifold. This brane world scenario, which incorporates the fixed points of the orbifold, gives a specific framework for modeling the extra dimensions of space beyond the familiar $4d$ Minkowski spacetime. The $4d$ branes exist at the fixed point of the orbifold. The bulk fields transform in the following way under the given parity transformations 
\begin{align}
	Z_2: & \Phi(x,y)\rightarrow \Phi(x,-y)=T_\phi[P_0]\Phi(x,y)\\
	Z_2': & \Phi(x,y')\rightarrow \Phi(x,-y')=T_\phi[P_1]\Phi(x,y').
\end{align}
Both the parity transformation matrices have the eigenvalues $\pm 1$, and after classifying the fields by the parity eigenvalues as $(\pm,\pm)$, they can be Fourier expanded as \cite{Hebecker:2001wq}
\begin{align}
	\Phi_{(++)}(x,y)\sim \sum_{n=0}^\infty \phi_{(++)}^{2n}(x)\cos \frac{2ny}{R}\\
	\Phi_{(+-)}(x,y)\sim \sum_{n=0}^\infty \phi_{(+-)}^{2n+1}(x)\cos  \frac{(2n+1)y}{R}\\
	\Phi_{(-+)}(x,y)\sim \sum_{n=0}^\infty \phi_{(-+)}^{2n+1}(x)\sin \frac{(2n+1)y}{R}\\
	\Phi_{(--)}(x,y)\sim \sum_{n=0}^\infty \phi_{(--)}^{2n+2}(x)\sin \frac{(2n+2)y}{R}.
\end{align}
The massless fields $\phi_{(++)}^0$ or the zero modes appear only in the components with even $(++)$ parities. Unless all components of the non-singlet field have common $Z_2$ parities, a {symmetry reduction} occurs upon compactification since zero modes are absent in fields with an odd parity \cite{Kawamura:1999nj,Kawamura:2008gh}. 
Thus, there is a reduction of the symmetry associated with the bulk field to the symmetry associated with the multiplet comprising of the component of the bulk having common even $Z_2$ partiy values. 
\section{Grand unification in $S^1/ Z_2 \times Z_2'$ orbifold}
\label{grand unification}
We consider the $5$-dimensional ($\mathcal{N}=2$) supersymmetric orbifold model following from \cite{Hebecker:2001wq}. We will also assume $R-$parity conservation in this model to suppress very fast $B,L$ violating processes \cite{Steven:2016ok}. In this model, the $5$-dimensional bulk fields consist of $SU(5)$ gauge super multiplets $\mathcal{V}$. In $4$-dimension $\mathcal{V}$ is decomposed as 
\begin{equation}
	\mathcal{V}=(V,\Sigma);\ \ V=\begin{pmatrix}
		A^\alpha_\mu(x,y)\\
		\lambda^\alpha(x,y)
	\end{pmatrix}^T;\ \Sigma= \begin{pmatrix}
		\sigma^\alpha(x,y)+iA^\alpha_5(x,y)\\
		\psi^\alpha(x,y)
	\end{pmatrix}^T
\end{equation}
$\alpha=1,2,...24$ denotes the $SU(5)$ generators. We again choose the fundamental representation matrices of $Z_2$ and $Z_2'$ as \cite{Hebecker:2001wq, Kawamura:2008gh}
\begin{equation}\label{Kr}
	P_0=\text{diag}(1,1,1,1,1)\quad\quad P_1= \text{diag}(-1,-1,-1,1,1).
\end{equation}
We have the following boundary conditions of the components of $\mathcal{N}=2$ vector multiplet for the $Z_2$ and $Z_2'$ transformations
\begin{align}
	V(x,y)&\xrightarrow[]{Z_2}V(x,-y)= P_0 V^\alpha_\mu(x,y)T^\alpha P_0^{\dag} \label{eq:p10}\\
	\Sigma(x,y)&\xrightarrow[]{Z_2}\Sigma(x,-y)= -P_0 \Sigma^\alpha_5(x,y)T^\alpha P_0^{\dag} \label{eq:p20}\\
	V (x,y')&\xrightarrow[]{Z_2'}A_\mu(x,-y')= P_1 V^\alpha_\mu(x,y)T^\alpha P_1^{\dag} \label{eq:p30}\\
	\Sigma (x,y')&\xrightarrow[]{Z_2'}\Sigma(x,-y')= -P_1\Sigma^\alpha_5(x,y)T^\alpha P_1^{\dag}. \label{eq:p40}
\end{align}
The positive or negative sign factors on the right-hand sides of eq.\eqref{eq:p10}-eq.\eqref{eq:p40} are the intrinsic
$Z_2$ parity, which is determined by the field strength and kinetic terms of the Lagrangian.  
$P_0,P_1$ commutes with standard model generators $T^a$, $a=1,...,12$ and thus only standard model $\mathcal{N}=1$ vector-supermultiplet has $(++)$ parity values \cite{Hebecker:2001wq, Kawamura:2008gh}, that is, only $V^a=(A^a_\mu, \lambda^a)$ remains unbroken after the orbifold breaking mechanism. The boundary conditions break the bulk symmetry to the Standard Model gauge group at $y=\pi R/2$ while the bulk symmetry remains intact at $y=0$ \cite{Hebecker:2001wq}.
The bulk fields $\mathcal{F}$ transform in the following way under the $SU(5)$ gauge group
\begin{equation}
\begin{split}
&\mathcal{F}\xrightarrow[]{SU(5)}U\mathcal{F},\\
&U=\text{exp}\left[i\xi_a(x,y)T^a+i\xi_{\hat{a}}(x,y)T^{\hat{a}}\right].    
\end{split}
\end{equation}
where, $a$ corresponds to standard model generators and $\hat{a}$ for the extra $SU(5)$ generators appearing in the theory. 
The transformation parameters $\xi_a,\xi_{\hat a}$ have the following mode expansion  \cite{Sarkar:2008xir}
\begin{align}
     \xi_a(x,y) = \sum_0^\infty \xi_a^n(x) \cos{2ny/R} \\
      \xi_{\hat a}(x,y) = \sum_0^\infty \xi_{\hat a}(x) \cos{(2n+1)y/R}.
\end{align}
Clearly, all the transformation parameters $\xi_{\hat a}$ vanishes at $y=\pi R/2$, and therefore at $y=0$ and $y\neq \pi R/2$, the bulk symmetry ($SU(5)$) is a good symmetry, while only the standard model gauge symmetry is remaining at the brane fixed at $y=\pi R/2$.
Hence we now add $SU(2)$ weak doublet chiral superfields $H_u, H_d$ on the boundary of $y=\pi R/2$. This solves the doublet-triplet splitting problem since we do not have the coloured Higgs triplet in the first place in our model.
\section{No proton decay in orbifold $SU(5)$ GUT}
\label{no pdecay}

We will now consider the fermions (quarks and leptons) as bulk fields and then demonstrate that they possess the necessary parity values to propagate on the brane fixed at $y=\pi R/2$.
The bulk obeys the whole $SU(5)$ symmetry thus, the usual $\textbf{10}+\overline{\textbf 5}$ representation is used for describing the matter content of one generation. We would obey the following convention regarding the parity transformations of the fermions (quarks and leptons)
\begin{align}
	P_i:q(x,y)&\xrightarrow[]{}\eta_i^q P_i q(x,y)\\
	P_i:q^c(x,y)&\xrightarrow[]{}-\eta_i^q P_i q^c(x,y)\\
	P_i:l(x,y)&\xrightarrow[]{}\eta_i^l P_i l(x,y)\\
	P_i:l^c(x,y)&\xrightarrow[]{}-\eta_i^l P_i l^c(x,y).
\end{align}
where $q^c,l\  (q,l^c)\in \overline{\textbf{5}}\ (\overline{\textbf{5}}^c)$. Similarly for the $q,q^c,l^c\in \textbf{10}$ we have 
\begin{align}
	P_i:q(x,y)&\xrightarrow[]{}\xi_i^q P_i q(x,y)P_i^{-1}\\
	P_i:q^c(x,y)&\xrightarrow[]{}-\xi_i^q P_i q^c(x,y)P_i^{-1}\\
	P_i:l(x,y)&\xrightarrow[]{}\xi_i^l P_i l(x,y)P_i^{-1}\\
	P_i:l^c(x,y)&\xrightarrow[]{}-\xi_i^l P_i l^c(x,y)P_i^{-1}.
\end{align}
In next fig., we draw proton decay at tree level by $X_{\mu}$ exchange. 
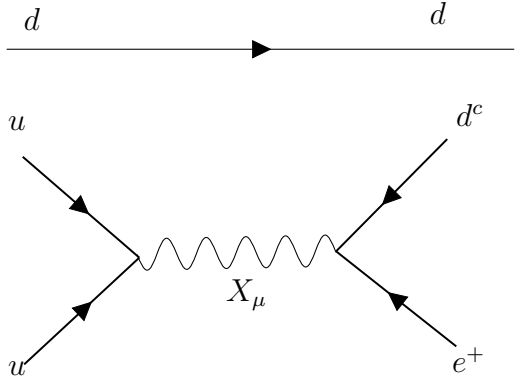
\begin{figure}[H]
\label{proton decay}
\centering
\begin{tikzpicture}[x=0.75pt,y=0.75pt,yscale=-1,xscale=1]

\draw   (248.27,160.64) .. controls (249.96,164.68) and (251.58,168.53) .. (253.39,168.49) .. controls (255.2,168.46) and (256.7,164.55) .. (258.27,160.44) .. controls (259.83,156.34) and (261.34,152.43) .. (263.14,152.39) .. controls (264.95,152.36) and (266.57,156.2) .. (268.26,160.24) .. controls (269.96,164.28) and (271.58,168.13) .. (273.38,168.09) .. controls (275.19,168.06) and (276.69,164.15) .. (278.26,160.04) .. controls (279.83,155.94) and (281.33,152.03) .. (283.14,151.99) .. controls (284.95,151.96) and (286.57,155.8) .. (288.26,159.84) .. controls (289.95,163.88) and (291.57,167.73) .. (293.38,167.69) .. controls (295.19,167.65) and (296.69,163.75) .. (298.26,159.64) .. controls (299.83,155.54) and (301.33,151.63) .. (303.14,151.59) .. controls (304.95,151.55) and (306.57,155.4) .. (308.26,159.44) .. controls (309.95,163.48) and (311.57,167.33) .. (313.38,167.29) .. controls (315.19,167.25) and (316.69,163.34) .. (318.26,159.24) .. controls (319.82,155.13) and (321.33,151.23) .. (323.13,151.19) .. controls (324.94,151.15) and (326.56,155) .. (328.25,159.04) .. controls (329.95,163.08) and (331.57,166.92) .. (333.37,166.89) .. controls (335.18,166.85) and (336.68,162.94) .. (338.25,158.84) .. controls (339.82,154.73) and (341.32,150.82) .. (343.13,150.79) .. controls (344.94,150.75) and (346.56,154.6) .. (348.25,158.64) .. controls (348.37,158.91) and (348.48,159.18) .. (348.59,159.45) ;
\draw    (182.5,57) -- (438.5,57) ;
\draw [shift={(315.5,57)}, rotate = 180] [fill={rgb, 255:red, 0; green, 0; blue, 0 }  ][line width=0.08]  [draw opacity=0] (8.93,-4.29) -- (0,0) -- (8.93,4.29) -- cycle    ;
\draw [color={rgb, 255:red, 0; green, 0; blue, 0 }  ,draw opacity=1 ][line width=0.75]    (249.41,162.19) -- (191.34,216.07) ;
\draw [shift={(225.14,184.71)}, rotate = 137.15] [fill={rgb, 255:red, 0; green, 0; blue, 0 }  ,fill opacity=1 ][line width=0.08]  [draw opacity=0] (8.93,-4.29) -- (0,0) -- (8.93,4.29) -- cycle    ;
\draw [color={rgb, 255:red, 0; green, 0; blue, 0 }  ,draw opacity=1 ][line width=0.75]    (190.61,111.28) -- (249.41,162.19) ;
\draw [shift={(223.79,140.01)}, rotate = 220.88] [fill={rgb, 255:red, 0; green, 0; blue, 0 }  ,fill opacity=1 ][line width=0.08]  [draw opacity=0] (8.93,-4.29) -- (0,0) -- (8.93,4.29) -- cycle    ;
\draw [color={rgb, 255:red, 0; green, 0; blue, 0 }  ,draw opacity=1 ][line width=0.75]    (348.66,158.99) -- (404.71,102.33) ;
\draw [shift={(372.12,135.28)}, rotate = 314.69] [fill={rgb, 255:red, 0; green, 0; blue, 0 }  ,fill opacity=1 ][line width=0.08]  [draw opacity=0] (8.93,-4.29) -- (0,0) -- (8.93,4.29) -- cycle    ;
\draw [color={rgb, 255:red, 0; green, 0; blue, 0 }  ,draw opacity=1 ][line width=0.75]    (409.28,206.99) -- (348.66,158.99) ;
\draw [shift={(375.05,179.89)}, rotate = 38.37] [fill={rgb, 255:red, 0; green, 0; blue, 0 }  ,fill opacity=1 ][line width=0.08]  [draw opacity=0] (8.93,-4.29) -- (0,0) -- (8.93,4.29) -- cycle    ;

\draw (405.93,206.05) node [anchor=north west][inner sep=0.75pt]    {$e^{+}$};
\draw (189.94,34.74) node [anchor=north west][inner sep=0.75pt]    {$d$};
\draw (292,171.86) node [anchor=north west][inner sep=0.75pt]    {$X_{\mu }$};
\draw (182,89.85) node [anchor=north west][inner sep=0.75pt]    {$u$};
\draw (408,83.24) node [anchor=north west][inner sep=0.75pt]    {$d^c$};
\draw (182,213.21) node [anchor=north west][inner sep=0.75pt]    {$u$};
\draw (394.74,31.74) node [anchor=north west][inner sep=0.75pt]    {$d$};

\end{tikzpicture}

\caption{Proton decay at tree level by $X_{\mu}$ exchange}
\end{figure}
Now, we have 
\[
P_0: \overline{\textbf{5}}(x,y)\xrightarrow[]{Z_2}\begin{pmatrix}
	-\eta_0^d d^c (x,y)\\
	\eta_0^e L_e(x,y)
\end{pmatrix}\quad 
P_1: \overline{\textbf{5}}(x,y')\xrightarrow[]{Z_2'}\begin{pmatrix}
	\eta_1^d d^c (x,y)\\
	\eta_1^e L_e(x,y)
\end{pmatrix}
\]
\[
P_0: \textbf{10}(x,y)\xrightarrow[]{Z_2}\begin{pmatrix}
	-\xi^u_0u^c(x,y) & -\xi^q_0 Q^\dag(x,y) \\
	\xi_0^q Q(x,y) & -\xi_0^e E^+(x,y)
\end{pmatrix}\quad 
P_1 :{\textbf{10}}(x,y')\xrightarrow[]{Z_2'}\begin{pmatrix}
	\xi^u_1u^c(x,y) & -\xi^q_1 Q^\dag(x,y) \\
	\xi_1^q Q(x,y) & \xi_1^e E^+(x,y)
\end{pmatrix}
\]
We choose 
\begin{equation}\label{ip}
	\eta_0^d=\eta_1^d=-\eta_0^e=-\eta_1^e=-1; \ \ \ \xi^u_0=-\xi_u^1=\xi_0^e=-\xi_1^e=-\xi_0^q=\xi_1^q=-1
\end{equation}
Finally, after considering eq.\eqref{ip}, we have the following parity assignments $(P_0,P_1)$ for all the particle content in $\overline{\textbf{5}}+\textbf{10}$ of $SU(5)$ 
\begin{align}
	L_e, \overline{U}, E^+ & \rightarrow  (+,+)\\
	d^c, Q  & \rightarrow  (+, -)\\
	L_e^c, \overline{U}^c, E & \rightarrow  (-,-)\\
	d, Q^c  & \rightarrow  (-, +).
\end{align}
\begin{table}[h]
\centering
\renewcommand{\arraystretch}{1.5} 
\begin{tabular}{|l|l|}
\hline
\textbf{Field Contents} & \textbf{Parity Assignments} \\
\hline
$L_e$, $\overline{U}$, $E^+$ & $(+,+)$ \\
\hline
$d^c$, $Q$ & $(+, -)$ \\
\hline
$L_e^c$, $\overline{U}^c$, $E$ & $(-,-)$ \\
\hline
$d$, $Q^c$ & $(-, +)$ \\
\hline
$(d^c)'$, $Q'$ & $(+, +)$ \\
\hline
$L_e'$, $\overline{U}'$, $(E^+)'$ & $(+,-)$ \\
\hline
$d'$, $(Q^c)'$ & $(-, -)$ \\
\hline
$(L_e^c)'$, $(\overline{U}^c)'$, $E'$ & $(-,+)$ \\
\hline
\end{tabular}
\caption{Field Contents and Parity Assignments}
\end{table}
Thus only ${\nu_e}_{L}, e, e^+, u^c$ in Standard Model have even $Z_2$ parities, and their zero modes survive in the 4-dimensional world. To get the other particles in the 4-dimensional brane after the orbifold breaking mechanism, we consider another copy of the $\textbf{10}+\overline{\textbf{5}}$ (denoted as ${\textbf{10}}'+\overline{\textbf{5}}'$) in the bulk and flip the sign of $P_1$, that is, $P_1'=\text{diag}(1,1,1,-1,-1)$. Now the parity assignments of the fields in $\textbf{10}'+\overline{\textbf{5}}'$ are 
\begin{align}
	(d^c)', Q'  & \rightarrow  (+, +)\\
	L_e', \overline{U}', (E^+)' & \rightarrow  (+,-)\\
	d', (Q^c)'  & \rightarrow  (-, -)\\
	(L_e^c)', (\overline{U}^c)', E' & \rightarrow  (-,+).
\end{align}
Combining the results obtained for the parity assignments of the particle content in both $\textbf{10}+\overline{\textbf{5}}$ and ${\textbf{10}}'+\overline{\textbf{5}}'$ representations, we find that all the Standard Model matter fields have zero modes, thus can be projected in the 4-dimensional wall \cite{Hebecker:2001wq}. \\\\
We summarise the field contents, and parity assignments in the table. 

\paragraph{Parities of gauge bosons.}
In the model, $4d$ gauge Field $A_{\mu}$ has to be even under $\mathbb{Z}_2$ parity operation because it has to be a symmetry of the theory 
\begin{equation}
	F_{\mu \nu}=\partial_{\mu}A_{\nu} - \partial_{\nu}A_{\mu}+i[A_{\mu},A{\nu}].
\end{equation}
$F_{\mu \nu}$ does not change covariantly under the parity operation if $A_{\mu}$ is odd under $ y\rightarrow -y$.
Now,
\begin{equation}
	F_{\mu5}=\partial_{\mu}A_{5}-\partial_{5}A_{\mu}+i[A_{\mu},A_5].
\end{equation} 
Under the parity operation $\partial_{\mu}$ does not change sign, $\partial_5$ changes sign under $ y\rightarrow -y$, so $A_5$ change sign under parity so the whole thing will be odd under parity operation. Since, the action includes $F^{\mu 5}F_{\mu 5}$ so $\mathcal{L}$ is invariant under the parity operation.
Therefore, $A_{\mu}$ and $A_{5}$ has $+$ and $-$ parities respectively under $y\rightarrow -y$.

With all Standard Model matter particles located at the brane fixed at \( y=\pi R/2 \), we can examine the potential for proton decay within the \( S^1/Z_2 \times Z_2' \) orbifold GUT model, where matter fields are distributed in bulk. As mentioned in the previous section, only $V^a=(A^a_\mu, \lambda^a)$ remains unbroken after the orbifold breaking mechanism, thus mediators responsible for proton decay, the \( X_\mu \) and \( Y_\mu \) bosons, which lack zero modes and even \( Z_2 \) parities, cannot exist on the 4-dimensional brane at $y=\pi R/2$. 
Note that, when the matter fields are considered as the bulk field, even though, there will be no proton-decay of the zero modes of the proton, the non-zero modes of the quarks in the bulk would still contribute to the proton decay in the bulk space excluding the boundary at $\pi R/2$. 
We hope to suppress such decay by the non-zero modes in the bulk, however, as of now, we consider that our matter fields are fixed at one of the boundaries.

The reason there is no proton decay in orbifold $SU(5)$ grand unified theories (GUTs) can be explained as follows. The non-SM gauge bosons responsible for proton decay have $(+ -)$ $Z_2$ parity values, and thus they have non-zero values at the $y=0$. If we consider the Standard Model matter (quarks and leptons) to be localized at a specific location, $y=\pi R/2$, then we find that proton decay is not possible in this model. This is because the massive gauge bosons that would mediate proton decay do not exist at the fixed (singular) point at $y=\pi R/2$, that is, on the brane where the SM fermions are localized. 

\section{$n-\Bar{n}$ oscillation in orbifold $SU(5)$ GUT}
\label{nnbar}
In this section, we study the possibility of $n-\Bar n$ oscillation within the $S^1/ Z_2\times Z_2'$ orbifold with a non-supersymmetric $SU(5)$ gauge group as the bulk symmetry. In our model we consider the following field contents obeying the bulk symmetry 
\begin{align*}
    H_{15} \in \textbf{15}\\
    H_{10} \in \textbf{10}\\
    \Sigma \in \textbf{24}.
\end{align*}
We fix the standard model matter (quarks and leptons) at $y=\pi R/2$, which is one of the fixed points of the orbifold. It should be noted that the representation $\textbf{15}$ can be decomposed as \cite{Sarkar:2008xir}
\begin{equation}
\textbf{15}=(\textbf{6},\textbf{1}, *)+(\textbf{3},\textbf{2}, *)+(\textbf{1},\textbf{3}, *).
\quad\quad (*\text{Hyper-charge to be insterted})
\end{equation}
In the bulk theory with $SU(5)$ gauge group, the unbroken Lagrangian term which would give rise to $d-d$ coupling is of the form \cite{Rao:1982gt}
\begin{equation}\label{eq:impshit}
	\mathcal{L}_{dd}=h \textbf{5}^T\ (H_{15}^{\dag})\textbf{5}
\end{equation}
with
\begin{equation}
\textbf{5}=\begin{pmatrix}
	d_1\\
	d_2\\
	d_3\\
	e^+\\
	\nu_e^c
\end{pmatrix}_R=\begin{pmatrix}
	\Vec{d}_R\\
	L_e^c
\end{pmatrix} ,\ 
~~H_{15} = \begin{pmatrix}
	\Delta_{3\times 3} & T_{3\times 2 }\\
	T^t_{2\times 3} & \lambda_{2\times 2}
\end{pmatrix}\ .
\end{equation}
The \textbf{10} of Higgs can be decomposed into 
$$\textbf{10}=(\overline{\textbf{3}},\textbf{1}, *)+(\textbf{3},\textbf{2}, *)+(\textbf{1},\textbf{3}, *),\quad\quad (*\text{Hyper-Charge to be inserted}).$$
The other required $SU(5)$ invariant Lagrangian term which would give rise to coupling of $u-u$ or $u-d$ is of the following form
\begin{equation}\label{umpa}
	\mathcal{L}_{uu/ud}=f \Sigma\  \overline{\textbf {10}} \ \textbf{10 }H_{10}.
\end{equation}
where
\[\textbf{10}=\begin{pmatrix}
    0 & {u}^c_g & -{u}^c_b & -u_r & -d_r\\
  -{u}^c_g  & 0&{u}^c_r &-u_b & -d_b\\
   {u}^c_b &-{u}^c_r & 0& -u_g&-d_g\\
   u_r &u_b &u_g &0 &-{e}^+\\
    d_r& d_b& d_g&{e}^+ &0
\end{pmatrix}= \begin{pmatrix}
      {U}_{3\times3}^c & -Q_{2\times3} \\
        Q_{3\times2}^t & E^+_{2\times2}
    \end{pmatrix},\ \ 
H_{10} = \begin{pmatrix}
	Y_{3\times 3} & \tau_{3\times 2 }\\
	-\tau^t_{2\times 3} & \Lambda_{2\times 2}
\end{pmatrix}.\  \ 
\]
Now, eq.\eqref{eq:impshit} can be written in terms of the matrix elements as 
\begin{equation}\label{eq:n-ns}
	h\left(\Vec{d}\Delta^\dag \Vec{d} + L_e^cT^\dag \Vec{d}+ \Vec{d}T^*L_e^c  + L_e^c\lambda^\dag L_e^c\right).
\end{equation}
Similarly, when $\Sigma$ gets it vev, eq.\eqref{umpa} can be written as 
\begin{equation}
\label{eq:nn}
    fv_0\begin{pmatrix}
2(UYU^c + Q^c \tau^T U^c + U \tau Q^T - Q^c \Lambda Q^T) & -2(UYQ + Q^c \tau^T Q + U \tau E^+ - Q^c \Lambda E^+) \\
-3({Q^c}^T Y U^c + E \tau^T U^c - {Q^c}^T \tau Q^T - E \Lambda Q^T) & 3({Q^c}^T Y Q - E \tau^T Q - {Q^c}^T \tau E^+ - E \Lambda E^+)
\end{pmatrix}
\end{equation}
Since we have fixed all the standard model particle content at the boundary $y=\pi R/2$, to check the feasibility of the interactions given in eq.\eqref{eq:n-ns} and eq.\eqref{eq:nn}, we have to find the components of $H_{15}$ and $H_{10}$ Higgs having positive $(+ +)$ $Z_2$ parity values. Only those interactions involving field content having  $(+ +)$ parity values would be possible in the four-dimensional world. Let us check the individual parities of the components of $H_{15}$ and $H_{10}$, 
\begin{equation}
  P_0: H_{15}\xrightarrow[]{Z_2}\begin{pmatrix}
	\eta_0^\Delta \Delta & \eta_0^T T\\
	\eta_0^T T^t & \eta_0^\lambda \lambda
\end{pmatrix},
\quad 
P_1 :H_{15}\xrightarrow[]{Z_2'}\begin{pmatrix}
	-\eta_1^\Delta \Delta & \eta_1^T T\\
	\eta_1^T T^t & -\eta_1^\lambda \lambda
\end{pmatrix},  
\end{equation}
\begin{equation}
P_0: H_{10}\xrightarrow[]{Z_2}\begin{pmatrix}
	\chi_0^Y Y & \chi_0^\tau \tau\\
	-\chi_0^\tau \tau^t & \chi_0^\Lambda \Lambda
\end{pmatrix},
\quad 
P_1: H_{10}\xrightarrow[]{Z_2'}\begin{pmatrix}
	-\chi_1^Y Y & \chi_1^\tau \tau\\
	-\chi_1^\tau \tau^t & -\chi_1^{\Lambda}\Lambda
\end{pmatrix}.   
\end{equation}
We choose $\eta_0^\Delta=\eta_0^T=-\eta_0^\lambda=1$, $\eta_1^\Delta=\eta_0^T=-\eta_0^\lambda=-1$, $\chi^Y_0=\chi^\tau_0=-\chi^\Lambda_0=1$ and $\chi^Y_1=\chi^\tau_1=-\chi^\Lambda_1=-1$ as the intrinsic parity values of the various components of the Higgs fields, then the various parity assignments corresponding to the all the field content are given in the table given below 
\begin{table}[h]
\centering
\renewcommand{\arraystretch}{1.5} 
\label{table1}
\begin{tabular}{|l|l|}
\hline
\textbf{Field Contents} & \textbf{Parity Assignments} \\
\hline
$\Delta$ & $(+,+)$ \\
\hline
$T$ & $(+,-)$ \\
\hline
$\lambda$ & $(-,+)$ \\
\hline
$Y$ & $(+,+)$ \\
\hline
$\tau$ & $(+,-)$ \\
\hline
$\Lambda$ & $(-,+)$ \\
\hline
\end{tabular}
\caption{Field Contents and Parity Assignments}
\end{table}\\
Therefore, only $h \Vec{d}\Delta^\dag \Vec{d}$, $2fv_0 UYU^c$, $-2fv_0 UYQ$, $-3fv_0 (Q^c)^T Y U^c$ and $3fv_0 Q^c Y Q$ from eq.\eqref{eq:n-ns}, and eq.\eqref{eq:nn} survive in the four-dimensional world. 

If we examine the Feynman diagram for neutron-antineutron oscillation as depicted in figure \ref{fig:nnbar}, the $d-d$ interaction is represented by the term $h \vec{d}\Delta^\dag \vec{d}$, while the $u-d$ interaction is represented by $-2fv_0 U Y Q$. In this context, we identify $X_1$ as $\Delta$ and $X_2$ as $Y$ in figure \ref{fig:nnbar}. The triscalar interaction shown in figure \ref{fig:nnbar} involves a term of the form $X_1 X_2 X_2$ or $\Delta Y Y$. This term should technically exist in the four-dimensional framework since both $\Delta$ and $Y$ have $(++)$ parity values.
\begin{figure}[H] 
\centering 
\begin{tikzpicture}[x=0.75pt,y=0.75pt,yscale=-1,xscale=1]

\draw [color={rgb, 255:red, 0; green, 0; blue, 0 }  ,draw opacity=1 ][line width=0.75]    (301.69,89) -- (253.45,32.18) ;
\draw [shift={(281.78,65.54)}, rotate = 229.67] [fill={rgb, 255:red, 0; green, 0; blue, 0 }  ,fill opacity=1 ][line width=0.08]  [draw opacity=0] (8.93,-4.29) -- (0,0) -- (8.93,4.29) -- cycle    ;
\draw [color={rgb, 255:red, 0; green, 0; blue, 0 }  ,draw opacity=1 ][line width=0.75]    (344.29,29.05) -- (301.69,89) ;
\draw [shift={(320.1,63.1)}, rotate = 305.39] [fill={rgb, 255:red, 0; green, 0; blue, 0 }  ,fill opacity=1 ][line width=0.08]  [draw opacity=0] (8.93,-4.29) -- (0,0) -- (8.93,4.29) -- cycle    ;
\draw [color={rgb, 255:red, 74; green, 144; blue, 226 }  ,draw opacity=1 ][line width=1.5]  [dash pattern={on 1.69pt off 2.76pt}]  (301.69,89) -- (301.5,163) ;
\draw [shift={(301.62,117.7)}, rotate = 90.15] [fill={rgb, 255:red, 74; green, 144; blue, 226 }  ,fill opacity=1 ][line width=0.08]  [draw opacity=0] (11.61,-5.58) -- (0,0) -- (11.61,5.58) -- cycle    ;
\draw [color={rgb, 255:red, 74; green, 144; blue, 226 }  ,draw opacity=1 ][line width=1.5]  [dash pattern={on 1.69pt off 2.76pt}]  (301.5,163) -- (251.5,218) ;
\draw [shift={(271.93,195.53)}, rotate = 312.27] [fill={rgb, 255:red, 74; green, 144; blue, 226 }  ,fill opacity=1 ][line width=0.08]  [draw opacity=0] (11.61,-5.58) -- (0,0) -- (11.61,5.58) -- cycle    ;
\draw [color={rgb, 255:red, 74; green, 144; blue, 226 }  ,draw opacity=1 ][line width=1.5]  [dash pattern={on 1.69pt off 2.76pt}]  (301.5,163) -- (347.5,217) ;
\draw [shift={(328.91,195.18)}, rotate = 229.57] [fill={rgb, 255:red, 74; green, 144; blue, 226 }  ,fill opacity=1 ][line width=0.08]  [draw opacity=0] (11.61,-5.58) -- (0,0) -- (11.61,5.58) -- cycle    ;
\draw    (251.5,218) -- (252.5,288) ;
\draw [shift={(251.91,246.5)}, rotate = 89.18] [fill={rgb, 255:red, 0; green, 0; blue, 0 }  ][line width=0.08]  [draw opacity=0] (8.93,-4.29) -- (0,0) -- (8.93,4.29) -- cycle    ;
\draw    (343.5,213) -- (344.5,283) ;
\draw [shift={(343.91,241.5)}, rotate = 89.18] [fill={rgb, 255:red, 0; green, 0; blue, 0 }  ][line width=0.08]  [draw opacity=0] (8.93,-4.29) -- (0,0) -- (8.93,4.29) -- cycle    ;
\draw [color={rgb, 255:red, 0; green, 0; blue, 0 }  ,draw opacity=1 ][line width=0.75]    (251.5,218) -- (203.26,161.18) ;
\draw [shift={(231.58,194.54)}, rotate = 229.67] [fill={rgb, 255:red, 0; green, 0; blue, 0 }  ,fill opacity=1 ][line width=0.08]  [draw opacity=0] (8.93,-4.29) -- (0,0) -- (8.93,4.29) -- cycle    ;
\draw [color={rgb, 255:red, 0; green, 0; blue, 0 }  ,draw opacity=1 ][line width=0.75]    (386.09,153.05) -- (343.5,213) ;
\draw [shift={(361.9,187.1)}, rotate = 305.39] [fill={rgb, 255:red, 0; green, 0; blue, 0 }  ,fill opacity=1 ][line width=0.08]  [draw opacity=0] (8.93,-4.29) -- (0,0) -- (8.93,4.29) -- cycle    ;

\draw (243.33,14.33) node [anchor=north west][inner sep=0.75pt]    {$d$};
\draw (342.96,10.34) node [anchor=north west][inner sep=0.75pt]    {$d$};
\draw (311,121.4) node [anchor=north west][inner sep=0.75pt]    {$X_{1}$};
\draw (253,163.4) node [anchor=north west][inner sep=0.75pt]    {$X_{2}$};
\draw (329,165.4) node [anchor=north west][inner sep=0.75pt]    {$X_{2}$};
\draw (351.33,241.4) node [anchor=north west][inner sep=0.75pt]    {$d$};
\draw (261.33,244.4) node [anchor=north west][inner sep=0.75pt]    {$d$};
\draw (193,138.74) node [anchor=north west][inner sep=0.75pt]    {$u$};
\draw (385,132.74) node [anchor=north west][inner sep=0.75pt]    {$u$};

\end{tikzpicture}
\caption{$n-\bar{n}$ oscillation}\label{fig:nnbar}
\begin{tikzpicture}[x=0.75pt,y=0.75pt,yscale=-1,xscale=1]

\draw [color={rgb, 255:red, 0; green, 0; blue, 0 }  ,draw opacity=1 ][line width=0.75]    (301.69,89) -- (253.45,32.18) ;
\draw [shift={(281.78,65.54)}, rotate = 229.67] [fill={rgb, 255:red, 0; green, 0; blue, 0 }  ,fill opacity=1 ][line width=0.08]  [draw opacity=0] (8.93,-4.29) -- (0,0) -- (8.93,4.29) -- cycle    ;
\draw [color={rgb, 255:red, 0; green, 0; blue, 0 }  ,draw opacity=1 ][line width=0.75]    (344.29,29.05) -- (301.69,89) ;
\draw [shift={(320.1,63.1)}, rotate = 305.39] [fill={rgb, 255:red, 0; green, 0; blue, 0 }  ,fill opacity=1 ][line width=0.08]  [draw opacity=0] (8.93,-4.29) -- (0,0) -- (8.93,4.29) -- cycle    ;
\draw [color={rgb, 255:red, 74; green, 144; blue, 226 }  ,draw opacity=1 ][line width=1.5]  [dash pattern={on 1.69pt off 2.76pt}]  (301.69,89) -- (301.5,163) ;
\draw [shift={(301.62,117.7)}, rotate = 90.15] [fill={rgb, 255:red, 74; green, 144; blue, 226 }  ,fill opacity=1 ][line width=0.08]  [draw opacity=0] (11.61,-5.58) -- (0,0) -- (11.61,5.58) -- cycle    ;
\draw [color={rgb, 255:red, 74; green, 144; blue, 226 }  ,draw opacity=1 ][line width=1.5]  [dash pattern={on 1.69pt off 2.76pt}]  (301.5,163) -- (251.5,218) ;
\draw [shift={(271.93,195.53)}, rotate = 312.27] [fill={rgb, 255:red, 74; green, 144; blue, 226 }  ,fill opacity=1 ][line width=0.08]  [draw opacity=0] (11.61,-5.58) -- (0,0) -- (11.61,5.58) -- cycle    ;
\draw [color={rgb, 255:red, 74; green, 144; blue, 226 }  ,draw opacity=1 ][line width=1.5]  [dash pattern={on 1.69pt off 2.76pt}]  (301.5,163) -- (347.5,217) ;
\draw [shift={(328.91,195.18)}, rotate = 229.57] [fill={rgb, 255:red, 74; green, 144; blue, 226 }  ,fill opacity=1 ][line width=0.08]  [draw opacity=0] (11.61,-5.58) -- (0,0) -- (11.61,5.58) -- cycle    ;
\draw    (251.5,218) -- (252.5,288) ;
\draw [shift={(251.91,246.5)}, rotate = 89.18] [fill={rgb, 255:red, 0; green, 0; blue, 0 }  ][line width=0.08]  [draw opacity=0] (8.93,-4.29) -- (0,0) -- (8.93,4.29) -- cycle    ;
\draw    (343.5,213) -- (344.5,283) ;
\draw [shift={(343.91,241.5)}, rotate = 89.18] [fill={rgb, 255:red, 0; green, 0; blue, 0 }  ][line width=0.08]  [draw opacity=0] (8.93,-4.29) -- (0,0) -- (8.93,4.29) -- cycle    ;
\draw [color={rgb, 255:red, 0; green, 0; blue, 0 }  ,draw opacity=1 ][line width=0.75]    (251.5,218) -- (203.26,161.18) ;
\draw [shift={(231.58,194.54)}, rotate = 229.67] [fill={rgb, 255:red, 0; green, 0; blue, 0 }  ,fill opacity=1 ][line width=0.08]  [draw opacity=0] (8.93,-4.29) -- (0,0) -- (8.93,4.29) -- cycle    ;
\draw [color={rgb, 255:red, 0; green, 0; blue, 0 }  ,draw opacity=1 ][line width=0.75]    (386.09,153.05) -- (343.5,213) ;
\draw [shift={(361.9,187.1)}, rotate = 305.39] [fill={rgb, 255:red, 0; green, 0; blue, 0 }  ,fill opacity=1 ][line width=0.08]  [draw opacity=0] (8.93,-4.29) -- (0,0) -- (8.93,4.29) -- cycle    ;

\draw (243.33,14.33) node [anchor=north west][inner sep=0.75pt]    {$u$};
\draw (342.96,10.34) node [anchor=north west][inner sep=0.75pt]    {$u$};
\draw (311,121.4) node [anchor=north west][inner sep=0.75pt]    {$X_{1}$};
\draw (253,163.4) node [anchor=north west][inner sep=0.75pt]    {$X_{2}$};
\draw (329,165.4) node [anchor=north west][inner sep=0.75pt]    {$X_{2}$};
\draw (351.33,241.4) node [anchor=north west][inner sep=0.75pt]    {$d$};
\draw (261.33,244.4) node [anchor=north west][inner sep=0.75pt]    {$d$};
\draw (193,138.74) node [anchor=north west][inner sep=0.75pt]    {$d$};
\draw (385,132.74) node [anchor=north west][inner sep=0.75pt]    {$d$};

\end{tikzpicture}
\caption{$n-\bar{n}$ oscillation}\label{fig:nnbar2}
\end{figure} 
This specific tri-scalar interaction is described by the following Lagrangian term
\begin{equation}\label{eq:tri-scal}
    \mathcal{L}_{\Delta YY}= \kappa \Sigma H_{10}^\dag H_{15} H_{10} .
\end{equation}
By selecting the $Z_2$ parity matrices as defined in equation \eqref{Kr}, and determining suitable intrinsic parity values for the components of $H_{10}$ and $H_{15}$, we establish all the interactions required for the $n-\Bar{n}$ oscillation within the $4$-dimensional universe, positioned at $y=\pi R/2$ as illustrated in figure \ref{fig:nnbar}.
If we consider the other quark-mixing interaction terms which can exist in the $4$-dimensional world, we will get other possible Feynman diagrams describing $n-\Bar{n}$ oscillation.
\begin{itemize}
    \item If we consider $h\vec d \Delta^\dag \vec d$, and $2fv_0 U Y U^c$, then we will have the triscalar term $X_1X_2X_2$ or $\Delta\Delta Y$, which one $dd\ (dd^c)$ coupling along with 2 $uu\ (uu^c)$ couplings as given in figure \ref{fig:nnbar2}. In this case, the tri-scalar Lagrangian term becomes 
    \begin{equation}\label{eq:tri-scal1}
    \mathcal{L}_{\Delta \Delta Y}= \kappa \Sigma H_{15}^\dag H_{10} H_{15} .
\end{equation}
    \item The terms $-2fv_0 UYQ$, $-3fv_0 (Q^c)^T Y U^c$ and $3fv_0 Q^c Y Q$ will give rise to both the Feynman diagrams given in figure \ref{fig:nnbar}, and figure \ref{fig:nnbar2}, with $X_1X_2X_2$ as $YYY$, that is, $X_1=X_2=Y$, and thus the corresponding tri-scalar Lagrangian term is 
    \begin{equation}\label{eq:tri-scal2}
    \mathcal{L}_{Y YY}= \kappa \epsilon^5 \Sigma H_{10}^\dag H_{10} H_{10},
\end{equation}
where $\epsilon^5$ is the 5-dimensional Levi-Civita tensor.
\end{itemize}
Note that the interactions needed for neutron-antineutron transition could also be mediated by the non-SM gauge bosons, namely, $X_\mu,\ Y_\mu$. But in the previous section, we have shown that these heavy gauge bosons would not exist in the 4-dimensional world where we have fixed all our SM matter content. Thus, $n-\overline{n}$ oscillation mediated by non-SM gauge bosons is forbidden in our model. \\

The interactions from eq.\eqref{eq:n-ns} and eq.\eqref{eq:nn} which are allowed in our model would also lead to other $B$ (or $B-L$) violating processes, which can also contribute to the generation of Baryon-asymmetry of the universe. We have to find the decay length of such processes to check the consistency of our model with current accepted values of the same.
\section{Conclusions}
\label{conclusions}
We find that in a $5$-dimensional space with the $5$th dimension being $S^1/Z_2\times Z_2'$ orbifold, and supersymmetric $SU(5)$ gauge symmetry, triplet-doublet splitting problem is realized along with the suppression of tree level proton decay. 

In the model, the no proton decay tree-level processes seems intrinsically tied to the orbifold construction, and this could explain why proton decay has not been observed experimentally so far. Proton decay might still occur through higher-dimensional non-renormalizable operators, but the suppression of these operators depends on the specifics of the model. However, by assigning a particular parity to the matter fields, proton decay can be entirely prohibited. While the idea of using discrete symmetry to prevent proton decay is not new \eg see \cite{Sakai:1981pk}, the physical origin of the relevant parity is especially clear and well-defined in our current framework. Also, in a non-supersymmetric framework, with the addition of the $\textbf{10, 15}$ of the Higgs, our model is consistent with the $n-\Bar{n}$ oscillation which is a $B-L$ violating process ($\Delta(B-L)=2$). The $n-\Bar{n}$ oscillation might be one of the generators of the baryogenesis. Further, the additional \textbf {10, 15} of Higgs can even explain the light masses of the neutrinos in non-supersymmetric $SU(5)$ with $S^1/Z_2\times Z_2'$ orbifold without the addition of right-handed singlet neutrino fields \eg see \cite{Chang:2003gv}. We note that the choice of intrinsic parities of various field contents may seem arbitrary, and this might be explained through some more fundamental theory in future \cite{Kawamura:2008gh}.

A new baryogenesis model is used to study \(n-\bar{n}\) oscillation at the TeV scale in \cite{Gu:2011ff}, and it would be helpful to link this with our work. Also, \cite{Gu:2007mi, Chakdar:2014ifa, Gu:2006dc} explores how baryon asymmetry can be explained via neutrinogenesis, and it would be worthwhile to examine this mechanism in more detail within the context of our work. 

\paragraph{Phenomenological implications and experimental prospects.}
A baryogenesis model studies \( n - \bar{n} \) oscillation at the TeV scale \cite{Gu:2011ff}. Studying this in our framework could explain how \( n - \bar{n} \) oscillation contributes to baryogenesis. Specifically, studying the conditions under which TeV-scale oscillations could arise within our SU(5) GUT model may help clarify the symmetries required for successful baryogenesis in this context. 

Also, studies in \cite{Gu:2007mi, Chakdar:2014ifa, Gu:2006dc} explore baryon asymmetry generation via neutrinogenesis, which introduces lepton number violation through heavy neutrino decay. This mechanism could be linked to our model by examining how neutrino mass generation and oscillation phenomena might influence baryon asymmetry in the early universe. A detailed analysis of the interaction between our framework and neutrinogenesis could enhance our understanding of how baryon asymmetry and \( n - \bar{n} \) oscillation mechanisms are interconnected. We hope to address this question in future work.

Further, our model predicts the scalar mediators needed for the $n-\overline{n
}$ oscillation. These mediators could be checked with various Baryon number violating operators in \cite{Jm:2012clt}. In \cite{MTa:2022d}, an orbifold model of 6 dimensions ($M^4\times(S^1/Z_2)\times(S^1/Z_2')$) have been used to show the feasibility of $n-\overline{n}$ oscillations at TeV scale, along with the suppression of proton decay. 


The effective operator giving rise to $n-\overline{n}$ oscillation is a \textit{six-quark} operator $QQQQQQ$ of \textit{mass dimension 9}, and thus a suppression factor of $M^{-5}_n$, where $M_n$ is the scale of new physics where we could probe $n-\overline{n}$ oscillation. The mean life of a free neutron calculated from the e transition probability (from $|n,0\rangle \rightarrow |\overline{n},t\rangle$) is $\sim 900$ seconds, which translates to $M_n \sim 500$ TeV \cite{Jm:2012clt,MTa:2022d}, which is much lower than the scale of proton decay and could be tested through the future collider experiments. Thus, studying $n-\overline{n}$
oscillations could give rise to interesting physics having been tested through experiments.

\begin{itemize}
	\item We emphasized the potential for testing our orbifold model in future collider experiments at the $\sim$TeV scale. This is shown using a $9$-dimensional effective field operator, which gives neutron-antineutron oscillations.
	
	\item \textbf{Scope of effective field-theoretic analysis} \\
	Our study opens the door to investigating neutron-antineutron oscillations, as well as other baryon and lepton number-violating processes, through two distinct effective field-theoretic approaches: 
	\begin{itemize}
		\item \textit{Bottom-Up Approach.} This approach involves constructing all possible Standard Model operators that could give rise to neutron-antineutron oscillations. By doing so, we aim to match the coefficients of these operators to the predictions of our orbifold model. 
		
		\item \textit{Top-Down Approach.} In this approach, we begin with new physics by constructing a Lagrangian containing heavier fields, such as scalar-bilinears, which mediate neutron-antineutron transitions linked to the GUT scale. These heavier fields are then integrated out, enabling their effects to be tested through low-energy standard model experiments.
	\end{itemize}
	
	According to \cite{Nusinov:B-X}, in an effective field theoretic approach, the bound on scale of new physics for an extra-dimensional model is given by 
    \[
    M_n  \gtrsim 45 \text{TeV} \times \left(\frac{\tau_{n\overline{n}}}{1.2\times 10^8 \text{sec}}  \right)^{1/9} \times \left(\frac{\mu}{3\times 10^3 \text{TeV}}\right)^{4/9}\times \left(\frac{|\langle \overline{n}|\mathcal{O}|n\rangle|}{10^{-4}\text{GeV}^6}
    \right)^{1/9}
    \]
    where the strongest bound on $n\rightarrow \overline{n}$ oscillation time is $\tau_{n\overline{n}}>2.6\times 10^8 \text{ sec}$ obtained by Super-Kamiokande for Oxygen 
\cite{Exp-NN}, \cite{Exp-NN2}. $\mu^{-1}<<\pi R/2$ is the half-width of the fermion wavefunction localization. The matrix element $\langle \overline{n} | \mathcal{O}| n \rangle$ was calculated to be around $\sim 10^{-4}\text{ GeV}^6$ in the MIT bag model for its two benchmark parameter values in \cite{Rao:1983sd}, where $\mathcal{O}$ are the low-energy 9-dimensional operators giving rise to $n-\overline{n}$ oscillations \cite{Rao:1982gt},\cite{Rao:1983sd}.

\item Furthermore, according to \cite{Sarkar:2008xir}, due to the effects of extra-dimensional modes in our GUT set-up, the unification scale would be around 30TeV,  these
are likely to be detected in the next generation
of collider experiments and various schemes are being planned to have high sensitive search for neutron-antineutron oscillation as given in \cite{Exp-NN}.

	\end{itemize}
The entire analysis done through effective operators will be the main focus of our next work utilizing 9-dimensional low-energy operators. According to \cite{Andrea-FOPT}, the matter-antimatter asymmetry could be probed by gravitational waves generated due to the first-order electroweak phase transitions because of $n-\overline{n}$ oscillations, and we plan to carefully analyze the Lagrangian terms of the form as given in \ref{eq:tri-scal1},\ref{eq:tri-scal2}, consider their zero-temperature 1-loop contributions, finite temperature thermal contributions, and test the possibility of strong first order (electro-weak) phase transition as in \cite{Andrea-FOPT}, calculate the bubble dynamics of the phase transition to find the gravitational wave spectrum and check if LISA, DECIGO, BBO, and other GW experiments could probe the gravitational wave produced in an orbifold set-up. 
 \section*{Acknowledgements}
The work of A.D. is supported by the SERB grant of the Indian Institute of Technology Kanpur. The work of S.D. is supported by the Shuimu Tsinghua Scholar Program of Tsinghua University. The work of U.S. is supported by a research grant associated with the Raja Ramanna
Fellowship of DAE, India.

\providecommand{\href}[2]{#2}\begingroup\raggedright\endgroup

\begin{thebibliography}{10}
\bibitem{Georgi:1974sy}
H.~Georgi and S.~L.~Glashow,
\emph{{Unity of All Elementary Particle Forces}},
\href{https://doi.org/10.1103/PhysRevLett.32.438}{\emph{Phys. Rev. Lett.} {\bfseries 32} (1974), 438-441}.

\bibitem{Pati:1973uk}
J.~C.~Pati and A.~Salam,
\emph{{Unified Lepton-Hadron Symmetry and a Gauge Theory of the Basic Interactions}},
\href{https://doi.org/10.1103/PhysRevD.8.1240}{\emph{Phys. Rev. D} \textbf{8} (1973) 1240-1251}.

\bibitem{Dimopoulos}
S.~Dimopoulos, S.~Raby, and F.~Wilczek,
\emph{{Supersymmetry and the scale of unification}},
Phys. Rev. D \textbf{24}, 1681.
\href{https://doi.org/10.1103/PhysRevD.24.1681}{\emph{Phys. Rev. D} \textbf{24} (1981) 1681-1683}.

\bibitem{Dimopoulos:1981zb}
S.~Dimopoulos and H.~Georgi,
\emph{{Softly Broken Supersymmetry and SU(5)}},
\href{https://doi.org/10.1016/0550-3213(81)90522-8}{\emph{Nucl. Phys. B} \textbf{193} (1981) 150-162}.

\bibitem{Hebecker:2001wq}
A.~Hebecker and J.~March-Russell,
\emph{{A Minimal $S^1/Z_2\times
Z_2'$ orbifold GUT}},
\href{https://doi.org/10.1016/S0550-3213(01)00374-1}{\emph{Nucl. Phys. B} \textbf{613} (2001) 3-16},
[\href{https://arxiv.org/abs/hep-ph/0106166}{{\ttfamily hep-ph/0106166}}].

\bibitem{Kawamura:1999nj}
Y.~Kawamura,
\emph{{Gauge symmetry breaking from extra space $S^1 / Z_2$
}}, \href{https://doi.org/10.1143/PTP.103.613}{\emph{Prog. Theor. Phys.} \textbf{103} (2000), 613-619},[\href{https://arxiv.org/abs/hep-ph/9902423}{{\ttfamily hep-ph/9902423}}].

\bibitem{Kawamura:2008gh}
Y.~Kawamura,
\emph{{Search for a Realistic Orbifold Grand Unification}},
\href{https://doi.org/10.1063/1.2939049}{\emph{AIP Conf. Proc.} \textbf{1015} (2000), (2008) no.1, 159-177},[\href{https://arxiv.org/abs/0802.3261}{{\ttfamily 0802.3261}}].

\bibitem{Rao:1982gt}
S.~Rao and R.~Shrock,
\emph{{$n \leftrightarrow \bar{n}$ Transition Operators and Their Matrix Elements in the {MIT} Bag Model}},
\href{https://doi.org/10.1016/0370-2693(82)90333-1}{\emph{Phys. Lett. B} \textbf{116} (2000), (1982), 238-242}.

\bibitem{Rao:1983sd}
S.~Rao and R.~E.~Shrock,
\emph{{Six Fermion ($B-L$) Violating Operators of Arbitrary Generational Structure}},
\href{https://doi.org/10.1016/0550-3213(84)90365-1}{\emph{Nucl. Phys. B} \textbf{232} (1984), 143-179}.

\bibitem{Kawamura:2000ev}
Y.~Kawamura,
\emph{Triplet doublet splitting, proton stability and extra dimension},
\href{https://doi.org/10.1143/PTP.105.999}{\emph{Prog. Theor. Phys.} \textbf{105} (2001), 999-1006},
[\href{https://arxiv.org/abs/hep-ph/0012125}{{\ttfamily hep-ph/0012125}}].

\bibitem{Altarelli:2001qj}
G.~Altarelli and F.~Feruglio,
\emph{SU(5) grand unification in extra dimensions and proton decay}, \href{https://doi.org/10.1016/S0370-2693(01)00650-5}{\emph{Phys. Lett. B} \textbf{511} (2001), 257-264},
[\href{https://arxiv.org/abs/hep-ph/0102301}{{\ttfamily hep-ph/0102301}}].

\bibitem{Hall:2001pg}
L.~J.~Hall and Y.~Nomura,
\href{https://doi.org/10.1103/PhysRevD.64.055003}{\emph{Phys. Rev. D} \textbf{64} (2001), 055003},
[\href{https://arxiv.org/abs/hep-ph/0103125}{{\ttfamily hep-ph/0103125}}].

\bibitem{Sarkar:2008xir}
U.~Sarkar,
\emph{Particle and astroparticle physics},
Taylor \& Francis, 2008,
ISBN 978-1-58488-931-1.

\bibitem{Steven:2016ok}
Stephen P. ~Martin,
\emph{A Supersymmetry Primer}
\href{https://arxiv.org/pdf/hep-ph/9709356}{9709356}

\bibitem{Chang:2003gv}
W.~F.~Chang and J.~N.~Ng,
\emph{Neutrino masses in 5-D orbifold SU(5) unification models without right handed singlets}, 
\href{https://doi.org/10.1088/1126-6708/2003/10/036}{\emph{JHEP} \textbf{10} (2003), 036}, [\href{https://arxiv.org/abs/hep-ph/0308187}{{\ttfamily hep-ph/0308187}}].

\bibitem{Sakai:1981pk}
N.~Sakai and T.~Yanagida,
\emph{Proton Decay in a Class of Supersymmetric Grand Unified Models},
\href{https://doi.org/10.1016/0550-3213(82)90457-6}{\emph{Nucl. Phys. B} \textbf{197} (1982), 533}.


\bibitem{Gu:2011ff}
P.~H.~Gu and U.~Sarkar,
\emph{Baryogenesis and neutron-antineutron oscillation at TeV},
\href{https://doi.org/10.1016/j.physletb.2011.10.017}{\emph{Phys. Lett. B} \textbf{705} (2011), 170-173}, [\href{https://arxiv.org/abs/1107.0173}{{\ttfamily 1107.0173}}].

\bibitem{Gu:2007mi}
P.~H.~Gu, H.~J.~He and U.~Sarkar,
\emph{Dirac neutrinos, dark energy and baryon asymmetry},
\href{https://doi.org/10.1088/1475-7516/2007/11/016}{\emph{JCAP} \textbf{11} (2007), 016}, 
[\href{https://arxiv.org/abs/0705.3736}{{\ttfamily 0705.3736}}].

\bibitem{Chakdar:2014ifa}
S.~Chakdar, K.~Ghosh and S.~Nandi,
\emph{A predictive model of Dirac Neutrinos},
\href{https://doi.org/10.1016/j.physletb.2014.05.036}{\emph{Phys. Lett. B} \textbf{734} (2014), 64-68}, [\href{https://arxiv.org/abs/1403.1544}{{\ttfamily 1403.1544}}].

\bibitem{Gu:2006dc}
P.~H.~Gu and H.~J.~He,
\emph{Neutrino Mass and Baryon Asymmetry from Dirac Seesaw},
\href{https://doi.org/10.1088/1475-7516/2006/12/010}{\emph{JCAP} \textbf{12} (2006), 010},
[\href{https://arxiv.org/abs/hep-ph/0610275}{{\ttfamily hep-ph/0610275}}].

\bibitem{Jm:2012clt}
J. ~M. ~Arnold, B. ~Fornal and M ~B. ~Wise,
\emph{Simplified models with baryon number violation but no proton decay
}
\href{https://journals.aps.org/prd/abstract/10.1103/PhysRevD.87.075004}{\emph{Phys. Rev. D} \textbf{87}(2013)}
[\href{https://arxiv.org/pdf/1212.4556}{{\ttfamily 1212.4556}}]

\bibitem{MTa:2022d}
M. ~T. ~Arun and D. ~Choudhury,
\emph{Neutron oscillation and baryogenesis from six dimensions}
\href{https://journals.aps.org/prd/abstract/10.1103/PhysRevD.106.L031701}{\emph{Phys. Rev. D \textbf{106}}(2022)}
[
\href{https://arxiv.org/pdf/2205.03846}{2205.03846}
]

\bibitem{Nusinov:B-X}
S. Nussinov and R. Shrock,
\emph{$n-\overline{n}$ Oscillations in Models with Large Extra Dimensions}

\bibitem{Andrea-FOPT}
A. Addazi,
\emph{Baryon violating first order phase transitions and gravitational waves},
\href{https://www.sciencedirect.com/science/article/pii/S0370269324006622}{\emph{Phys. Lett. B \textbf{859}}(2024)}

\bibitem{Exp-NN}
A. Addazi, et. al.,
\emph{New high-sensitivity searches for neutrons converting
into antineutrons and/or sterile neutrons at the European
Spallation Source}
\href{https://iopscience.iop.org/article/10.1088/1361-6471/abf429}{\emph{ J. Phys. G: Nucl. Part. Phys. \textbf{48} }(2021)}
[\href{https://arxiv.org/pdf/2006.04907}{2006.0490}]

\bibitem{Exp-NN2}
K. Abe, et. al.,
\emph{Search for $n-\overline{n}$
 oscillation in Super-Kamiokande}
 \href{https://journals.aps.org/prd/abstract/10.1103/PhysRevD.91.072006}{\emph{Phys. Rev. D \textbf{91}}(2015)}
 [\href{https://arxiv.org/pdf/1109.4227}{1109.4227}]
\end{thebibliography}
\end{document}